\documentclass[aps,prd,a4paper,twocolumn,amsmath,nofootinbib,preprintnumbers,notitlepage]{revtex4-1}

\pdfoutput=1
\usepackage{amssymb,amsmath,latexsym,mathrsfs}
\usepackage[sort&compress]{natbib}
\usepackage{graphicx,subfigure}
\usepackage{epsfig}
\usepackage{varioref,xr-hyper}
\usepackage{color}
\usepackage{multirow}
\usepackage{array}
\usepackage{hyperref}
\usepackage{wasysym}
\usepackage{color}
\usepackage{float}
\usepackage{xcolor}
\usepackage[utf8]{inputenc}
\usepackage[T1]{fontenc}

\begin{document}



\title{Emergent Dark Energy, neutrinos and cosmological tensions}

\author{Weiqiang Yang}
\email{d11102004@163.com}
\affiliation{Department of Physics, Liaoning Normal University, Dalian, 116029, People's Republic of China}

\author{Eleonora Di Valentino}
\email{eleonora.divalentino@manchester.ac.uk}
\affiliation{Jodrell Bank Center for Astrophysics, School of Physics and Astronomy, 
University of Manchester, Oxford Road, Manchester, M13 9PL, UK.}

\author{Supriya Pan}
\email{supriya.maths@presiuniv.ac.in}
\affiliation{Department of Mathematics, Presidency University, 86/1 College Street, Kolkata 700073, India}

\author{Olga Mena}
\email{omena@ific.uv.es }
\affiliation{IFIC, Universidad de Valencia-CSIC, 46071, Valencia, Spain}

\begin{abstract}
The Phenomenologically Emergent Dark Energy model, a dark energy model with the same number of free parameters as the flat $\Lambda$CDM, has been proposed as a working example of a minimal model which can avoid the current cosmological tensions. A straightforward question is whether or not the inclusion of massive neutrinos and extra relativistic species may spoil such an appealing phenomenological alternative. We present the bounds on $M_{\nu}$ and $N_{\rm eff}$ and comment on the long standing $H_0$ and $\sigma_8$ tensions within this cosmological framework with a wealth of cosmological observations. Interestingly, we find, at $95\%$ confidence level, and with the most complete set of cosmological observations, $M_{\nu}\sim 0.21^{+0.15}_{-0.14}$ eV  and $N_{\rm eff}= 3.03\pm 0.32$ i.e. an indication for a non-zero neutrino mass with a significance above $2\sigma$. The well known Hubble constant tension is considerably easened, with a significance always below the $2\sigma$ level. 
\end{abstract}

\maketitle
\section{Introduction}

Despite the success of the standard cosmological $\Lambda$CDM paradigm to explain the current observations, our picture of the universe, and more concretely, of its current accelerated expansion, is far to be complete. There are a number of unsolved questions and pending inconsistencies that need to be addressed. For example, the simplest cosmological scenario based on  General Relativity together with a positive cosmological constant as the dark energy component faces the well-known $H_0$ and $S_8$ tensions\footnote{The parameter $S_8$ is proportional to $\sigma_8 \sqrt{\Omega_m}$, with $\sigma_8$ indicating  the amplitude  of  the  small-scale  density  fluctuations.}, which have motivated plenty of searches for new physics scenarios, ranging from the nature of dark matter and/or dark radiation~\cite{Dvorkin:2014lea,Berezhiani:2015yta,Bernal:2016gxb,Ko:2016uft,Kumar:2016zpg,Buen-Abad:2017gxg,Ko:2016fcd,Chacko:2016kgg,Gariazzo:2017pzb,Feng:2017nss,Zhao:2017urm,DEramo:2018vss,Alexander:2019rsc,Escudero:2019gvw,Vattis:2019efj,Alcaniz:2019kah,Blinov:2020hmc,Clark:2020miy,Archidiacono:2020yey,Hryczuk:2020jhi} dark energy~\cite{DiValentino:2015ola,Poulin:2018cxd,Vagnozzi:2019ezj,DiValentino:2016hlg,Qing-Guo:2016ykt,Karwal:2016vyq,Zhao:2017cud,DiValentino:2017zyq,Raveri:2017jto,DiValentino:2017oaw,DiValentino:2017rcr,Sola:2017znb,Mortsell:2018mfj,Banihashemi:2018has,Yang:2018qmz,Poulin:2018zxs,Sakstein:2019fmf,Niedermann:2019olb,Yang:2019jwn,Shafieloo:2016bpk,Li:2019san,Yang:2020zuk,Li:2020ybr,DiValentino:2020naf,DiValentino:2020kha,Chudaykin:2020acu,Alestas:2020mvb,Niedermann:2020dwg}, exotic dark matter-dark energy interactions~\cite{Xia:2016vnp,DiValentino:2017iww,Kumar:2017dnp,Yang:2018uae,Yang:2019uzo,Kumar:2019wfs,DiValentino:2019dzu,Pan:2019jqh,Martinelli:2019dau,DiValentino:2019ffd,DiValentino:2019jae,Yang:2018euj,Pan:2019gop,Yang:2017ccc,Pan:2020bur}, modified gravity~\cite{Dirian:2017pwp,Renk:2017rzu,Khosravi:2017hfi,Peirone:2017vcq,Nunes:2018xbm,Yan:2019gbw,Raveri:2019mxg,Cai:2019bdh,Frusciante:2019puu,Wang:2020zfv}, and many others~\cite{Kreisch:2019yzn,Banihashemi:2018oxo,Colgain:2018wgk,Guo:2018ans,Agrawal:2019lmo,Arendse:2019hev,Garcia-Quintero:2019cgt,Hart:2019dxi,Liu:2019awo,DiValentino:2019exe,Yang:2019qza,Desmond:2019ygn,Ivanov:2019hqk,Colgain:2019joh,Visinelli:2019qqu,Berghaus:2019cls,DiValentino:2020hov,Jedamzik:2020krr,Pan:2020zza,Wu:2020nxz,Ye:2020btb,Ballesteros:2020sik,Braglia:2020iik,Ballardini:2020iws,Keeley:2020rmo,Capozziello:2020nyq,Gonzalez:2020fdy}, most of them involving extensions of the parameter space compared to the $\Lambda$CDM model.

If we go beyond the simplest $\Lambda$CDM scenario, some of the problems associated to the standard cosmology may be solved. Unfortunately, some new complications naturally arise. On the other hand, (assuming the spatial flatness of the Friedmann-Lema\^{i}tre-Robertson-Walker (FLRW) geometry), the number of free parameters in the $\Lambda$-cosmology is only six. Any extension of the $\Lambda$-cosmology in terms of new degrees freedom naturally exacerbates  the $\chi^2$ of the corresponding observational analysis compared to the $\Lambda$-cosmology.  Therefore, the construction of a \emph{minimal} cosmological model is very important.  The modifications could also rely on the gravitational sector of the theory \cite{PhysRev.124.925,Banerjee:2000gt,Banerjee:2000mj,Capozziello:2002rd,Das:2005bn,Nojiri:2006ri,Sotiriou:2008rp,DeFelice:2010aj,Nojiri:2010wj,Capozziello:2011et,deHaro:2012zt,Cai:2015emx,Paliathanasis:2015aos,Paliathanasis:2016vsw,Nunes:2016qyp,Nunes:2016drj,Paliathanasis:2017efk,Dimakis:2017kwx,Nojiri:2017ncd,Nunes:2018evm,Paliathanasis:2019luv,Das:2018ylw}, rather than in the stress-energy tensor  \cite{Copeland:2006wr,Basilakos:2011rx,Paliathanasis:2014yfa,Dimakis:2016mip,Basilakos:2018xjp,Dimakis:2019qfs,Papagiannopoulos:2019kar,Banerjee:2005ef,Banerjee:2005vy,Banerjee:2006na,deHaro:2016hpl,deHaro:2016hsh,deHaro:2016ftq,deHaro:2016cdm,Pan:2016ngu,Sharov:2017iue,Pan:2017ios,Pan:2017zoh,Yang:2018xah,Yang:2017yme,Pan:2017ent,Yang:2018qec,Pan:2019brc,Pan:2020mst}. 

With such a wide model building perspective in mind, the authors of Ref.~\cite{Li:2019yem} proposed a possible alternative framework, the Phenomenologically Emergent Dark Energy (PEDE) model\footnote{Perhaps the readers might be interested in a recent work on the Emergent Dark Energy having field theoretic origin \cite{Barker:2020elg}. However, in this work we shall focus on the PEDE model proposed in \cite{Li:2019yem} and recently extended with one extra parameter in the dark sector in Ref.~\cite {Li:2020ybr}.}.
This scenario has exactly the same number of free parameters as in the spatially flat $\Lambda$-cosmology and it has been shown to provide a solution to the $H_0$ tension. The very same model was investigated further in Ref.~\cite{Pan:2019hac} (see also Ref.~\cite{Rezaei:2020mrj}) considering its complete evolution including both the cosmological background plus perturbations and analyzed using the most recent  cosmological observations, obtaining similar results concerning the $H_0$ tension. Of course the success of the PEDE model in principle may rely on the different ingredients included in our Universe's description, and the nice features of this model may disappear if one starts to include extra parameters such as the dark matter or dark radiation  properties. Thus, the motivation of this article is to check how the extra degrees of freedom in terms of the neutrino properties could affect the constraints on the Hubble constant within this model scenario. Actually, it is well known that these additional parameters can potentially solve the sound horizon problem~\cite{Knox:2019rjx,Arendse:2019hev}, improving the agreement with the BAO data, contrarily to other parameters as, for example, the spatial curvature.
According to the observational data, 
neutrino oscillations have robustly established the existence of neutrino masses and their impact in cosmology is crucial (see e.g. Refs.~\cite{Lesgourgues:2006nd,Lattanzi:2017ubx,Aker:2019uuj}).

In order to assess the viability of the PEDE model, we include the total neutrino mass ($M_{\nu}$) as well as the effective number of neutrino species ($N_{\rm eff}$) in the minimal PEDE scenario explored throughout Ref.~\cite{Liu:2020vgn} and consider three  extended cosmological schemes, namely, PEDE $+ M_{\nu}$, PEDE $+ N_{\rm eff}$ and PEDE $+ M_{\nu}+ N_{\rm eff}$.

The manuscript is organized as follows. Section \ref{sec-2} introduces the PEDE model. In Sec.~\ref{sec-data} we describe the cosmological probes used to examine the models. After that, in Sec.~\ref{sec-results}, we discuss the results for all the analyzed models to conclude in Sec.~\ref{sec-conclu} with our main findings.

\section{Phenomenologically Emergent Dark Energy}
\label{sec-2}
 
Starting from a spatially flat Friedmann-Lema\^{i}tre-Robertson-Walker (FLRW)  line element, $ds^2  =  -dt^2 + a^2 (t) (dx^2 + dy^2 +dz^2)$, where $a(t)$ is the expansion scale factor of the universe and $(t, x, y, z)$ are the co-moving coordinates, and assuming that the matter sector is minimally coupled to gravity (described by the Einstein gravity), the gravitational field equations can be written as 
\begin{eqnarray}
H^2  = \frac{8 \pi G}{3} \left( \rho_{\nu} + \rho_r + \rho_m +\rho_{DE} \right); \label{F1}\\
2 \dot{H} + 3 H^2 = - 8 \pi G \left(p_{\nu} + p_r + p_m +p_{DE} \right),\label{F2}
\end{eqnarray}
where $G$ is the Newton's gravitational constant, the dot represents the time derivative, $H \equiv \dot{a}/a$ is the Hubble rate of the FLRW universe; $(\rho_i, p_i)$ [$i = \nu, r, m, {\rm DE}$] are respectively the energy density and pressure of  massive neutrinos, radiation, pressure-less matter (baryons+pressure-less dark matter) and the dark energy fluid. 
All the components of the stress energy tensor are assumed to be of barotropic nature and show no interactions among themselves. The conservation equation for each fluid is therefore $\dot{\rho}_i + 3 H (p_i +\rho_i) = 0$. If the dark energy fluid has a time-dependent cosmological constant, then using the conservation equation for DE, one can easily show that the energy density of DE (with respect to the critical energy density) defined as  $\Omega_{DE} \equiv \rho_{\rm DE}/\rho_{crit,0}$ evolves with the cosmic time (or alternatively with respect to the redshift $z$) as:

\begin{eqnarray}
\tilde{\Omega}_{DE} (z) = \Omega_{DE,0} \; \exp \left[ 3 \; \int_{0}^{z} \frac{1+w_{DE}(z^\prime)}{1+z^\prime} dz^\prime \right]~,
\label{eq:main}
\end{eqnarray}
where $\Omega_{\rm DE,0}$ is the present value of $\tilde{\Omega}_{\rm DE}$ and $w_{\rm DE}(z) = \frac{p_{\rm DE}}{\rho_{\rm DE}}$ is the equation-of-state of the dark energy fluid. Usually, by prescribing the equation-of-state for DE, one can determine the evolution of the DE density. Alternatively, one may apply the reverse engineering method by prescribing the parametric form for $\tilde{\Omega}_{\rm DE}$ and consequently determine other cosmological parameters. The Phenomenological Dark Energy 
Emergent model follows this reverse mechanism~\cite{Li:2019yem} and use as 
cosmic time the number of \emph{ten-foldings} of the scale factor\footnote{The scale factor and the redshift are related as $a\equiv \frac{1}{1+z}$.}, proposing a phenomenological description of the dark energy fluid as follows:
\begin{eqnarray}\label{model}
\tilde{\Omega}_{DE} (z) = \Omega_{DE,0} \Bigl[1- \tanh (\log_{10} (1+z)) \Bigr]~.
\end{eqnarray}
It is interesting to note that the PEDE model defined above has no extra free parameters, i.e. it has exactly the same number of free parameters as the flat $\Lambda$CDM model and also only one describing the dark energy fluid properties: $\Omega_{DE,0}$, the current dark energy energy density. Using the conservation equation for the DE and using Eq.~(\ref{eq:main}), the dark energy equation of state can be derived as:
\begin{eqnarray}\label{model}
w_{DE} (z) = \frac{1}{3}\frac {d\ln \tilde{\Omega}_{DE}}{dz} (1+z) -1~.
\end{eqnarray}

The authors of Ref.~\cite{Li:2019yem}  have shown that the dark energy equation of state in the PEDE model at early times would be  $-1 -2/(3\ln 10)$ to evolve asymptotically to $-1$ in the far future.  The current value is $w_{DE} (z=0) =-1 -1/(3\ln 10)$. 

Similarly to the work carried out in Ref.~\cite{Pan:2019hac}, we have also taken into account the evolution at the level of perturbations. In the next section we describe the cosmological datasets and the statistical methods to constrain the different cosmological scenarios.

\section{Cosmological data and Methodology}
\label{sec-data}

In order to analyze the PEDE framework in the presence of massive neutrinos and extra relativistic species, we have used a wealth of cosmological observations that we shall detail in what follows.

\begin{enumerate}

\item \textbf{Cosmic Microwave Background (CMB)}: We use the latest CMB data from the Planck 2018 legacy release~\cite{Aghanim:2018eyx,Aghanim:2019ame}\footnote{Concretely, we use the CMB temperature and polarization angular power spectra {\it plikTTTEEE+lowl+lowE} of~\cite{Aghanim:2018eyx,Aghanim:2019ame}}. 

\item \textbf{CMB Lensing:} We include the Planck 2018 CMB lensing reconstruction power spectrum data~\cite{Aghanim:2018oex}.

\item \textbf{Baryon acoustic oscillation (BAO) data}:  We use BAO data from various galaxy surveys that include  6dFGS~\cite{Beutler:2011hx}, SDSS-MGS~\cite{Ross:2014qpa}, and 
BOSS DR12~\cite{Alam:2016hwk}, as 
considered by the Planck 2018 data analyses~\cite{Aghanim:2018eyx}.

\item \textbf{Hubble constant (R19)}: The measurement of the Hubble constant yielding $H_0 = 74.03 \pm 1.42$ km/s/Mpc at $68\%$ CL by Riess et al. \cite{Riess:2019cxk} has been incorporated into our analysis. This measurement shows a $4.4 \sigma$ tension with that extracted from the CMB within the minimal $\Lambda$CDM model.

\item \textbf{Supernovae Type Ia}: we exploit measurements of SNIa luminosity distances from the Pantheon sample~\cite{Scolnic:2017caz}, comprising 1048 data points spanned over the redshift interval $z \in [0.01, 2.3]$.

\item \textbf{Dark energy survey (DES):} We use the galaxy clustering and cosmic shear measurements from DES combined-probe Year 1 results~\cite{Troxel:2017xyo, Abbott:2017wau, Krause:2017ekm}, as adopted by the Planck collaboration in their 2018 final data analyses~\cite{Aghanim:2018eyx}.

\end{enumerate}

For the analyses, we shall use a modified version of the publicly available Markov Chain Monte Carlo (MCMC) code \texttt{CosmoMC}~\cite{Lewis:2002ah,Lewis:1999bs} package (see \url{http://cosmologist.info/cosmomc/}). This code supports the new 2018 Planck likelihood~\cite{Aghanim:2019ame}, implements an efficient sampling of the posterior distribution using the fast/slow parameter decorrelations \cite{Lewis:2013hha}, and has a convergence diagnostic based on the Gelman-Rubin statistics~\cite{Gelman:1992zz}.  
Finally, we have computed the Bayesian Evidence (BE) for all the scenarios considered in this work. 
We have actually followed the \texttt{MCEvidence} code  ~\cite{Heavens:2017hkr,Heavens:2017afc}, where it has been shown that the BE can directly be computed using only the MCMC chains which are used to constrain the parameter space. A concise discussion on BE can be found in \cite{Yang:2018euj}. So, using the  \texttt{MCEvidence} code we have computed the logarithm of the Bayes factor, $B_{ij}$ of the cosmological scenarios with respect to the base $\Lambda$CDM model. That means we calculate $\ln B_{ij}$ (where $i=$ PEDE, or PEDE + $M_{\nu}$, PEDE + $N_{\rm eff}$, or PEDE + $M_{\nu}$ + $N_{\rm eff}$ and $j=\Lambda$CDM), and after that we use the revised Jeffreys scale shown in Table~\ref{tab:jeffreys} in order to quantify the  strength of evidence of the cosmological scenarios. 

\begingroup                                                                                                                     
\begin{center}                                                                                                                  
\begin{table}[!h]                                                                                                                
\begin{tabular}{cc}                                                                                                            
\hline\hline                                                                                                                    
$\ln B_{ij}$ & Strength of evidence for model $M_i$ \\ \hline
$0 \leq \ln B_{ij} < 1$ & Weak \\
$1 \leq \ln B_{ij} < 3$ & Definite/Positive \\
$3 \leq \ln B_{ij} < 5$ & Strong \\
$\ln B_{ij} \geq 5$ & Very strong \\
\hline\hline                                                                                                                    
\end{tabular}                                                                                                                   
\caption{Revised Jeffreys scale which quantifies the strength of evidence of the proposed cosmological model (identified through the index $i$) with respect to the base cosmological model $\Lambda$CDM (identified through the symbol $j$).  }\label{tab:jeffreys}                                                                                                   
\end{table}
\end{center}                                                                                                                    
\endgroup 
\section{Results}
\label{sec-results}

In this Section we first discuss the results obtained within the minimal PEDE scenario. Then, we enlarge the model to account for the neutrino parameters. We shall consider three different cases: a first one where we allow for the total neutrino mass $M_{\nu}$ to freely vary, a second one where we introduce the possibility of having an extra dark radiation component parameterized by $N_{\rm eff}$, and finally when we consider the previous two neutrino parameters to freely vary simultaneously.

\subsection{PEDE}

We begin the analysis with the simple and minimal PEDE model, updating the earlier work of Ref.~\cite{Pan:2019hac} by means of Planck 2018 data. The results are summarized in Tab.~\ref{tab:PEDE}. 

Comparing the new constraints with those one in Table II of Ref.~\cite{Pan:2019hac}, we notice that there is no such significant shifts on the cosmological parameters (with the exception of $\tau$ and $ln(10^{10}A_s)$ due to the new low-$\ell$ polarization data), and the conclusions about the PEDE model solving the Hubble constant tension still hold. Indeed, having the same degrees of freedom of the $\Lambda$CDM scenario, the PEDE model can strongly relieve the $H_0$ tension, leading it to negligible statistically significant levels: for instance, the tension is reduced to the $1\sigma$ level for Planck CMB data alone. Also $S_8\equiv \sigma_8 \sqrt{\Omega_m/0.3}$ shifts now towards smaller values, showing a better agreement with cosmic shear measurements. The fact that the $H_0$ tension is strongly alleviated in the PEDE scenario is due to the fact that the dark energy equation of state within this phenomenological picture is negative,  $w_{DE} (z=0) =-1/(3\ln 10)-1 \simeq -1.145$, as it is well known that a phantom dark energy equation of state can raise the value of $H_0$~\cite{DiValentino:2016hlg,DiValentino:2019jae}, especially when dealing with CMB data only. When $w$ is allowed to lie in the phantom region, the parameter $H_0$ must be increased to leave unchanged the location of the CMB acoustic peaks. The addition of BAO measurements mildly softens this degeneracy.

\begingroup                                                                                                                     
\squeezetable                                                                                                                   
\begin{center}                                                                                                                  
\begin{table*}                          \resizebox{\textwidth}{!}{                                                                                     
\begin{tabular}{cccccccccccc}                                                                                                            
\hline\hline                                                                                                                    
Parameters & Planck 2018 & Planck 2018+BAO & Planck 2018+R19 & Planck 2018+BAO+R19 & Full \\ \hline

$\Omega_c h^2$ & $    0.1200_{-    0.0015-    0.0025}^{+    0.0013+    0.0028}$ & $    0.1215_{-    0.0010-    0.0019}^{+    0.0010+    0.0020}$  & $    0.1193_{-    0.0012-    0.0023}^{+    0.0012+    0.0023}$ & $    0.12093_{-    0.00092-    0.0019}^{+    0.00096+    0.0019}$ & $    0.12032_{-    0.00079-    0.0015}^{+    0.00077+    0.0016}$  \\

$\Omega_b h^2$ & $    0.02238_{-    0.00015-    0.00030}^{+    0.00015+    0.00029}$ & $    0.02227_{-    0.00013-    0.00025}^{+    0.00013+    0.00026}$ & $    0.02243_{-    0.00014-    0.00027}^{+    0.00014+    0.00029}$ & $    0.02232_{-    0.00013-    0.00025}^{+    0.00013+    0.00026}$ & $    0.02236_{-    0.00013-    0.00024}^{+    0.00012+    0.00025}$ \\

$100\theta_{MC}$ & $    1.04092_{-    0.00031-    0.00062}^{+    0.00032+    0.00062}$ & $    1.04076_{-    0.00030-    0.00056}^{+    0.00030+    0.00057}$  & $    1.04102_{-    0.00031-    0.00062}^{+    0.00030+    0.00063}$ & $    1.04082_{-    0.00031-    0.00056}^{+    0.00028+    0.00060}$ & $    1.04085_{-    0.00031-    0.00055}^{+    0.00028+    0.00059}$ \\

$\tau$ & $    0.0544_{-    0.0080-    0.016}^{+    0.0080+    0.016}$ & $    0.0527_{-    0.0078-    0.015}^{+    0.0073+    0.016}$ & $    0.0555_{-    0.0077-    0.016}^{+    0.0080+    0.016}$ & $    0.05279_{-    0.0070-    0.014}^{+    0.0073+    0.015}$ & $    0.04756_{-    0.0070-    0.015}^{+    0.0077+    0.014}$ \\

$n_s$ & $    0.9651_{-    0.0044-    0.0084}^{+    0.0044+    0.0081}$ & $    0.9617_{-    0.0038-    0.0075}^{+    0.0038+    0.0074}$ & $    0.9670_{-    0.0040-    0.0081}^{+    0.0040+    0.0081}$ & $    0.9632_{-    0.0041-    0.0073}^{+    0.0037+    0.0076}$ & $    0.9634_{-    0.0037-    0.0070}^{+    0.0036+    0.0072}$ \\

${\rm{ln}}(10^{10} A_s)$ & $    3.044_{-    0.018-    0.033}^{+    0.016+    0.034}$ & $    3.044_{-    0.017-    0.030}^{+    0.015+    0.032}$  & $    3.045_{-    0.016-    0.034}^{+    0.016+    0.032}$ & $    3.043_{-    0.015-    0.029}^{+    0.015+    0.029}$ & $    3.030_{-    0.014-    0.030}^{+    0.014+    0.030}$  \\

$\Omega_{m0}$ & $    0.2734_{-    0.0092-    0.016}^{+    0.0078+    0.017}$ & $    0.2822_{-    0.0060-    0.012}^{+    0.0061+    0.012}$ & $    0.2691_{-    0.0068-    0.014}^{+    0.0068+    0.013}$ & $    0.2788_{-    0.0055-    0.011}^{+    0.0056+    0.011}$ & $    0.2753_{-    0.0049-    0.0088}^{+    0.0046+    0.0093}$  \\

$\sigma_8$ & $    0.8579_{-    0.0078-    0.015}^{+    0.0079+    0.016}$ & $    0.8612_{-    0.0076-    0.014}^{+    0.0070+    0.014}$  & $    0.8566_{-    0.0076-    0.016}^{+    0.0076+    0.015}$ & $    0.8597_{-    0.0068-    0.013}^{+    0.0068+    0.014}$ & $    0.8522_{-    0.0059-    0.013}^{+    0.0061+    0.013}$ \\

$H_0$ & $   72.35_{-    0.79-    1.5}^{+    0.78+    1.5}$ & $   71.54_{-    0.55-    1.1}^{+    0.53+    1.1}$  & $   72.76_{-    0.67-    1.3}^{+    0.65+    1.3}$ & $   71.84_{-    0.52-    1.0}^{+    0.51+    1.1}$ & $   72.16_{-    0.44-    0.87}^{+    0.44+    0.86}$ \\

$S_8$ & $    0.819_{-    0.018-    0.030}^{+    0.016+    0.032}$ & $    0.835_{-    0.013-    0.025}^{+    0.013+    0.023}$  & $    0.811_{-    0.015-    0.029}^{+    0.014+    0.028}$ & $    0.829_{-    0.012-    0.024}^{+    0.012+    0.023}$ & $    0.8163_{-    0.0089-    0.018}^{+    0.0091+    0.017}$ \\

$r_{\rm{drag}}$ & $  147.084_{-    0.285-    0.608}^{+    0.327+    0.557}$ & $  146.816_{-    0.235-    0.473}^{+    0.260+    0.457}$  & $  147.206_{-    0.269-    0.534}^{+    0.271+    0.540}$  & $  146.903_{-    0.241-    0.449}^{+    0.241+    0.471}$ & $  147.022_{-    0.203-    0.396}^{+    0.208+    0.419}$\\
\hline\hline                                                                                                                    
\end{tabular}  }                                                                                                                 
\caption{Observational constraints on the simple PEDE scenario at 68\% and 95\% CL using various cosmological datasets. The full combination refers to the datasets Planck 2018+BAO+R19+Pantheon+DES+Lensing. }
\label{tab:PEDE}                                                                                                   
\end{table*}                                                                                                                     
\end{center}                                                                                                                    
\endgroup                                                                                                                       

\subsection{PEDE $+M_{\nu}$}
\begin{figure*}
\includegraphics[width=0.5\textwidth]{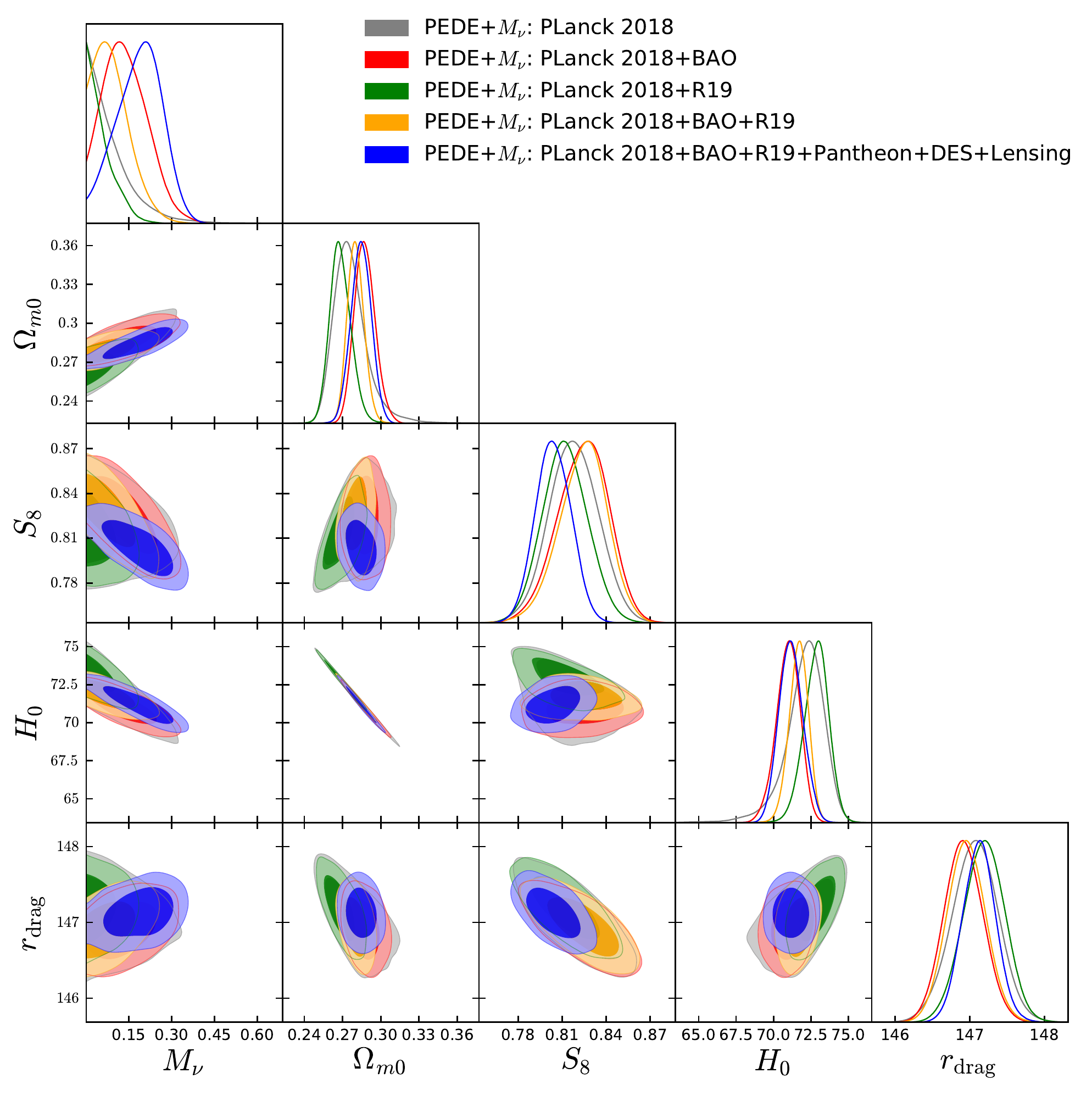}
\caption{One-dimensional marginalized posterior distributions for some selected parameters and two dimensional $68\%$ and $95\%$~CL joint contours for several combinations of the parameters within the extended scenario PEDE+$M_{\nu}$ using various cosmological datasets.}
\label{fig1}
\end{figure*}

We report the constraints at 68\% and 95\% CL on the cosmological parameters for the PEDE $+M_{\nu}$ model in Tab.~\ref{tab:mnu} and we show the 1D posterior distributions and 2D correlation plots in Fig.~\ref{fig1}.

Comparing Tab.~\ref{tab:mnu} with Tab.~\ref{tab:PEDE} it is evident that including a total neutrino mass free to vary does not change the constraints on the six free cosmological parameters of the model, showing the robustness of the model. However, both $H_0$ and $S_8$ are affected by the introduction of $M_{\nu}$, because of their important negative correlations, as can be noticed from Fig.~\ref{fig1}. A freely varying $M_{\nu}$ will imply a slightly lower value for the Hubble constant and $S_8$. This correlation is also very well known within the $\Lambda$CDM model, preventing the possibility of combining the Hubble constant direct measurements with the CMB because shifting down the $H_0$ mean value exacerbates further their tension. In the PEDE model, however, even if there is a slight shift of the mean value of $H_0$ when introducing $M_{\nu}$ as a free parameter, there is an increase of the corresponding error bars. Consequently, the Hubble constant value reported by Planck 2018 in the PEDE scenario is in agreement within 1$\sigma$ with $H_0$ measurements from R19 also when massive neutrinos are present, allowing us to combine the two measurements, with the aim of breaking the $M_{\nu}$ versus $H_0$ correlation and obtain more stringent constraints. Therefore, while we have at 95\% CL for Planck 2018 alone $M_{\nu}<0.26$ eV, we find for Planck 2018+R19 $M_{\nu}<0.15$~eV. Contrarily to the $\Lambda$CDM scenario, the addition of BAO data does not improve the upper limit at 95\% CL on the total neutrino mass ($M_{\nu}<0.27$ eV for Planck 2018+BAO), and this is due to the appearance of a peak for $M_{\nu}$ different from zero in the 1D posterior distribution (see the red curve in Fig.~\ref{fig1}). We find, at 68\% CL, $M_{\nu}=0.139^{+0.065}_{-0.093}$~eV.

When considering all the data combined, i.e.~Planck 2018+BAO+Pantheon+R19+DES+lensing, the Hubble constant tension persists albeit at the (mild) 1.8$\sigma$ level, and the $S_8$ parameter is shifted towards a lower value. Finally, we have an indication at more than $2\sigma$ for a total neutrino mass different from zero, i.e~$M_{\nu}=0.19^{+0.14}_{-0.16}$~eV. The fact that the neutrino mass bound is larger in the PEDE scenario is due, again, to the strong degeneracy between the dark energy equation of state and $M_{\nu}$. It is well-known that the neutrino mass bounds are less constraining within the phantom region~\cite{Vagnozzi:2018jhn,Choudhury:2018byy}, because in this region the normalized expansion rate is smaller than within the $\Lambda$CDM picture and therefore one needs to increase $M_{\nu}$ and/or $H_0$ to compensate for such an effect. 
Concerning the cosmic shear tension, it is also alleviated, as a larger neutrino mass implies a lower value of $S_8$, as can be noticed from Fig.~\ref{fig1}.

\begingroup                                                                                                                     
\squeezetable                                                                                                                   
\begin{center}                                                                                                                  
\begin{table*}                             \resizebox{\textwidth}{!}{                                                                                      
\begin{tabular}{ccccccccccccc}                                                                                                            
\hline\hline                                                                                                                    
Parameters & Planck 2018 & Planck 2018+BAO & Planck 2018+R19 & Planck 2018+BAO+R19 & Full \\ \hline
$\Omega_c h^2$ & $    0.1201_{-    0.0014-    0.0027}^{+    0.0014+    0.0028}$ &  $    0.1210_{-    0.0011-    0.0023}^{+    0.0012+    0.0021}$  & $    0.1194_{-    0.0012-    0.0023}^{+    0.0012+    0.0023}$  & $    0.1207_{-    0.0011-    0.0022}^{+    0.0011+    0.0021}$ & $    0.11969_{-    0.00086-    0.0017}^{+    0.00084+    0.0017}$  \\

$\Omega_b h^2$ & $    0.02237_{-    0.00015-    0.00030}^{+    0.00015+    0.00030}$ &  $    0.02229_{-    0.00014-    0.00026}^{+    0.00013+    0.00026}$  & $    0.02244_{-    0.00014-    0.00027}^{+    0.00014+    0.00028}$  & $    0.02233_{-    0.00013-    0.00027}^{+    0.00013+    0.00026}$  & $    0.02240_{-    0.00013-    0.00025}^{+    0.00013+    0.00025}$  \\

$100\theta_{MC}$ & $    1.04090_{-    0.00032-    0.00062}^{+    0.00031+    0.00063}$ & $    1.04075_{-    0.00029-    0.00060}^{+    0.00030+    0.00058}$  & $    1.04102_{-    0.00029-    0.00059}^{+    0.00029+    0.00060}$   & $    1.04085_{-    0.00029-    0.00058}^{+    0.00029+    0.00057}$  & $    1.04087_{-    0.00028-    0.00058}^{+    0.00028+    0.00055}$  \\

$\tau$ & $    0.0543_{-    0.0082-    0.015}^{+    0.0074+    0.016}$ &  $    0.0534_{-    0.0082-    0.015}^{+    0.0073+    0.016}$  & $    0.0544_{-    0.0079-    0.014}^{+    0.0072+    0.015}$  & $    0.0532_{-    0.0074-    0.015}^{+    0.0072+    0.016}$  & $    0.0535_{-    0.0090-    0.015}^{+    0.0076+    0.017}$ \\

$n_s$ & $    0.9651_{-    0.0044-    0.0087}^{+    0.0045+    0.0089}$ &  $    0.9628_{-    0.0038-    0.0073}^{+    0.0039+    0.0078}$  & $    0.9667_{-    0.0041-    0.0081}^{+    0.0041+    0.0083}$  & $    0.9636_{-    0.0039-    0.0075}^{+    0.0039+    0.0074}$ & $    0.9649_{-    0.0035-    0.0073}^{+    0.0039+    0.0071}$ \\

${\rm{ln}}(10^{10} A_s)$ & $    3.044_{-    0.016-    0.030}^{+    0.015+    0.032}$ &  $    3.044_{-    0.017-    0.031}^{+    0.015+    0.033}$   & $    3.043_{-    0.015-    0.030}^{+    0.015+    0.031}$  & $    3.044_{-    0.015-    0.030}^{+    0.016+    0.032}$  & $    3.042_{-    0.016-    0.031}^{+    0.016+    0.033}$\\

$\Omega_{m0}$ & $  0.277_{-    0.015-    0.025}^{+    0.009+    0.028}$ & $    0.2875_{-    0.0088-    0.016}^{+    0.0076+    0.017}$  & $    0.2684_{-    0.0090-    0.016}^{+    0.0077+    0.017}$ & $    0.2804_{-    0.0071-    0.013}^{+    0.0061+    0.013}$  & $    0.2846_{-    0.0079-    0.015}^{+    0.0078+    0.015}$ \\

$\sigma_8$ & $    0.852_{-    0.010-    0.045}^{+    0.024+    0.034}$ &  $    0.842_{-    0.017-    0.042}^{+    0.026+    0.039}$   & $    0.859_{-    0.010-    0.029}^{+    0.016+    0.025}$  & $    0.853_{-    0.013-    0.037}^{+    0.022+    0.033}$  & $    0.826_{-    0.021-    0.034}^{+    0.018+    0.036}$ \\

$H_0 {\rm [km/s/Mpc]}$ & $   72.1_{-    0.9-    2.8}^{+    1.5+    2.5}$ & $   70.98_{-    0.75-    1.6}^{+    0.85+    1.5}$  & $   72.85_{-    0.77-    1.7}^{+    0.89+    1.6}$  & $   71.68_{-    0.60-    1.3}^{+    0.70+    1.2}$   & $   71.17_{-    0.86-    1.5}^{+    0.76+    1.6}$  \\

$M_{\nu} {\rm [eV]}$ & $    <0.103\,,<0.26$ & $    0.139_{-    0.093}^{+    0.065} <0.27$  & $<0.067\,,<0.15$  & $    0.094_{-    0.088}^{+    0.029}<0.21$  & $    0.187_{-    0.073-    0.16}^{+    0.090+    0.14}$ \\

$S_8$ & $    0.818_{-    0.016-    0.033}^{+    0.017+    0.033}$ &  $    0.824_{-    0.017-    0.035}^{+    0.019+    0.034}$  & $    0.812_{-    0.015-    0.029}^{+    0.015+    0.030}$  & $    0.824_{-    0.015-    0.034}^{+    0.018+    0.033}$  & $    0.804_{-    0.012-    0.023}^{+    0.012+    0.023}$\\

$r_{\rm{drag}}$ & $  147.068_{-    0.304-    0.621}^{+    0.305+    0.601}$ & $  146.917_{-    0.260-    0.503}^{+    0.261+    0.521}$ &  $  147.190_{-    0.272-    0.548}^{+    0.276+    0.527}$  & $  146.955_{-    0.256-    0.507}^{+    0.255+    0.515}$ & $  147.124_{-    0.225-    0.432}^{+    0.219+    0.432}$ \\


\hline\hline                                                                                                                    
\end{tabular}    }                                                                                                               
\caption{Observational constraints at 68\% and 95\% CL using various cosmological datasets within the extended  PEDE $+ M_{\nu}$ scenario.}
\label{tab:mnu}                                                                                                   
\end{table*}                                                                                                                     
\end{center}                                                                                                                    
\endgroup                                                                                                                        

\subsection{PEDE $+N_{\rm eff}$}

\begin{figure*}
\includegraphics[width=0.5\textwidth]{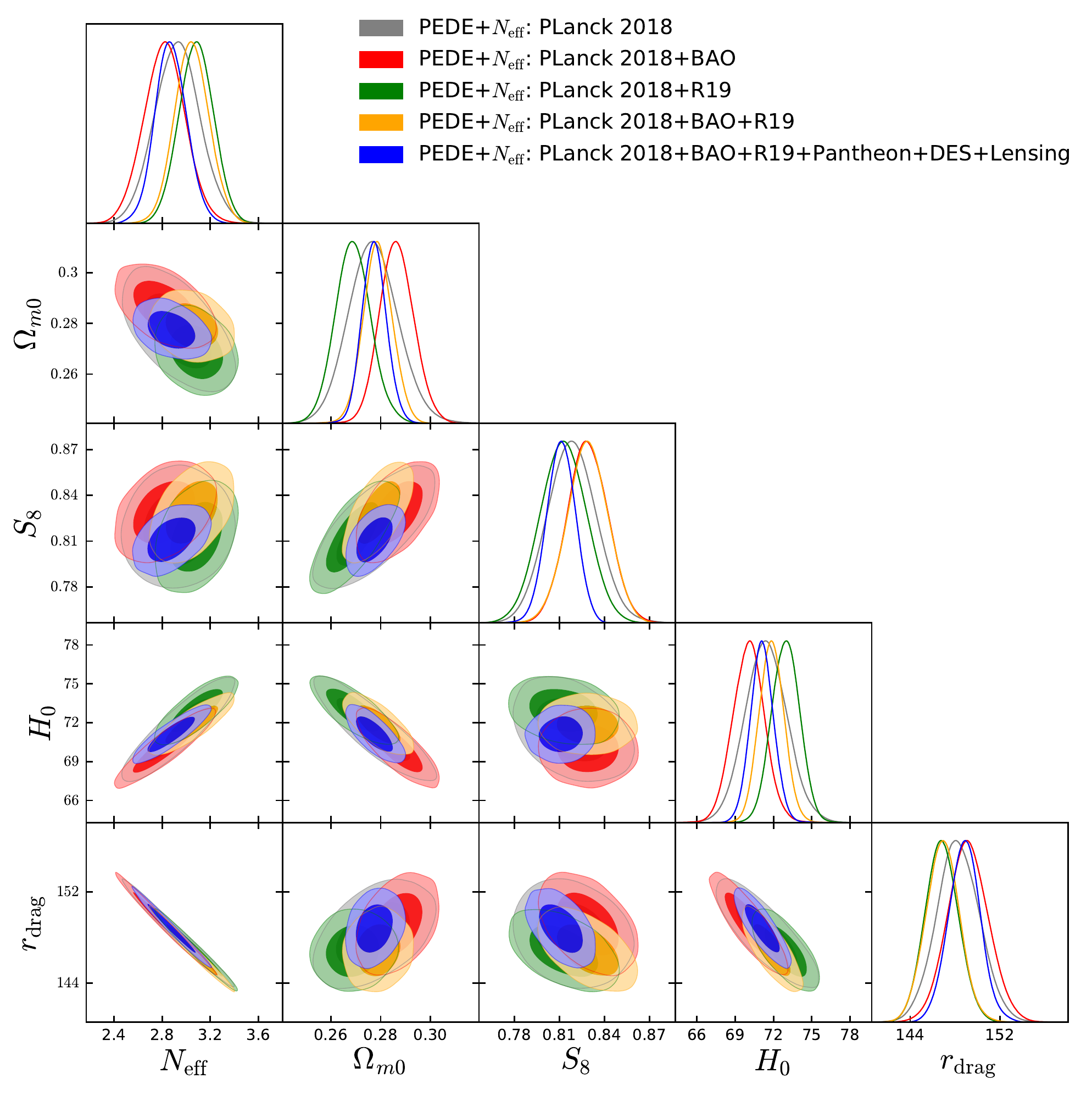}
\caption{One-dimensional marginalized posterior distributions for some selected parameters and two dimensional $68\%$ and $95\%$~CL joint contours for several combinations of the parameters within the extended scenario PEDE+$N_{\rm eff}$ using various cosmological datasets. }
\label{fig2}
\end{figure*}

We report the constraints at 68\% and 95\% CL on the cosmological parameters for PEDE $+N_{\rm eff}$ model in Tab.~\ref{tab:neff} and we show the 1D posterior distributions and 2D correlation plots in Fig.~\ref{fig2}.

Comparing Tab.~\ref{tab:neff} with Tab.~\ref{tab:PEDE} it is evident that including a number of relativistic species at recombination $N_{\rm eff}$ free to vary introduces the already known correlation between the cosmological parameters present in the $\Lambda$CDM scenario. Therefore, we have an increase of the error bars, and, since the $N_{\rm eff}$ mean value is, for all the combination of data, with the exception of Planck 2018+R19, lower than the expected value $3.045$~\cite{Mangano:2005cc,deSalas:2016ztq,Akita:2020szl}, we have a shift down in the value of the parameters positively correlated with $N_{\rm eff}$, as $\Omega_c h^2$, $n_s$ and $H_0$, and a shift up on the value of those negatively correlated, such as $\theta_{MC}$.

Contrarily to what happens within the $\Lambda$CDM $+N_{\rm eff}$ scenario, in the PEDE $+N_{\rm eff}$ model we have Planck 2018 in agreement with R19 at 1.2$\sigma$, justifying their combination.
The value of $N_{\rm eff}$ will be shifted from $N_{\rm eff}=2.93^{+0.18}_{-0.19}$ at 68\% CL for Planck 2018 (slightly low but perfectly consistent with the expected value) to $N_{\rm eff}=3.08\pm 0.14$ at 68\% CL for Planck 2018+R19.

For the Planck 2018+BAO dataset combination we have instead $N_{\rm eff}=2.82\pm0.17$ at 68\% CL, i.e. 1$\sigma$ lower than the expected value, and this is reflected on the shift of the Hubble constant parameter, raising the $H_0$ tension to 2.1$\sigma$. 
For the Planck 2018+BAO+R19 combination we again obtain a slightly larger $N_{\rm eff}=3.05\pm 0.14$ at 68\% CL due to the larger value of the Hubble constant when the R19 prior is considered in the data analyses.

Finally, in the full dataset combination case, i.e.~Planck 2018+BAO+Pantheon+R19+DES+lensing, we find that the Hubble constant tension appears with a modest significance (1.7$\sigma$). We also find a very robust constraint on the effective number of relativistic species, $N_{\rm eff}=2.87^{+0.13}_{-0.14}$ at 68\% CL, which is in perfect agreement with the expectation of the Standard Model for three neutrino families. Therefore, contrarily to what happens in the $\Lambda$CDM $+N_{\rm eff}$ scenario, in the PEDE $+N_{\rm eff}$ model a larger value of the Hubble constant in agreement with R19 does not necessarily imply an extra amount of dark radiation at recombination. This is again due to the fact that the larger value of $H_0$ in this model is strongly related to the fact that currently $w_{\rm DE}<-1$, becoming the dark energy equation-of-state more negative ($w_{\rm DE}\simeq-1.34$) as we move back to early times.

\begingroup                                                                                                                     
\squeezetable 
\begin{center}                                                                                                                  
\begin{table*}                              \resizebox{\textwidth}{!}{                                                                                     
\begin{tabular}{ccccccc}                                                                                                            
\hline\hline                                                                                                                    
Parameters & Planck 2018 & Planck 2018+BAO & Planck 2018+R19 & Planck 2018+BAO+R19  & Full \\ \hline
$\Omega_c h^2$ & $    0.1183_{-    0.0030-    0.0059}^{+    0.0030+    0.0061}$ & $    0.1178_{-    0.0030-    0.0058}^{+    0.0029+    0.0059}$  & $    0.2000_{-    0.0027-    0.0053}^{+    0.0027+    0.0053}$  & $    0.1210_{-    0.0026-    0.0053}^{+    0.0027+    0.0055}$  & $    0.1174_{-    0.0025-    0.0047}^{+    0.0023+    0.0048}$  \\

$\Omega_b h^2$ & $    0.02227_{-    0.00022-    0.00043}^{+    0.00022+    0.00044}$ & $    0.02211_{-    0.00018-    0.00036}^{+    0.00018+    0.00036}$   & $    0.02245_{-    0.00017-    0.00033}^{+    0.00017+    0.00032}$  & $    0.02233_{-    0.00016-    0.00031}^{+    0.00016+    0.00031}$  & $    0.02225_{-    0.00015-    0.00030}^{+    0.00015+    0.00030}$ \\

$100\theta_{MC}$ & $    1.04112_{-    0.00043-    0.00084}^{+    0.00044+    0.00085}$ & $    1.04118_{-    0.00045-    0.00086}^{+    0.00044+    0.00090}$  & $    1.04095_{-    0.00041-    0.00080}^{+    0.00041+    0.00080}$  & $    1.04085_{-    0.00041-    0.00079}^{+    0.00042+    0.00081}$  & $    1.04120_{-    0.00040-    0.00076}^{+    0.00039+    0.00078}$  \\

$\tau$ & $    0.0535_{-    0.0073-    0.015}^{+    0.0074+    0.015}$ & $    0.0515_{-    0.0072-    0.015}^{+    0.0073+    0.015}$  & $    0.0552_{-    0.0076-    0.016}^{+    0.0078+    0.016}$  & $    0.0528_{-    0.0075-    0.015}^{+    0.0074+    0.016}$  & $    0.0483_{-    0.0067-    0.014}^{+    0.0068+    0.014}$  \\

$n_s$ & $    0.9605_{-    0.0085-    0.016}^{+    0.0084+    0.017}$ & $    0.9541_{-    0.0068-    0.014}^{+    0.0068+    0.014}$  & $    0.9679_{-    0.0061-    0.012}^{+    0.0059+    0.012}$ & $    0.9630_{-    0.0059-    0.011}^{+    0.0054+    0.011}$  & $    0.9581_{-    0.0054-    0.011}^{+    0.0054+    0.011}$  \\

${\rm{ln}}(10^{10} A_s)$ & $    3.038_{-    0.018-    0.036}^{+    0.018+    0.036}$ & $    3.032_{-    0.018-    0.034}^{+    0.018+    0.036}$  & $    3.046_{-    0.017-    0.034}^{+    0.017+    0.035}$   & $    3.044_{-    0.017-    0.034}^{+    0.017+    0.033}$  & $    3.024_{-    0.014-    0.028}^{+    0.014+    0.029}$  \\

$\Omega_{m0}$ & $    0.2772_{-    0.0098-    0.019}^{+    0.0097+    0.020}$ & $    0.2865_{-    0.0073-    0.013}^{+    0.0065+    0.014}$  & $    0.2690_{-    0.0070-    0.014}^{+    0.0070+    0.015}$  & $    0.2789_{-    0.0057-    0.011}^{+    0.0058+    0.012}$  & $    0.2775_{-    0.0051-    0.0091}^{+    0.0047+    0.010}$  \\

$\sigma_8$ & $    0.852_{-    0.012-    0.024}^{+    0.012+    0.024}$ & $    0.848_{-    0.012-    0.023}^{+    0.012+    0.024}$  & $    0.859_{-    0.011-    0.022}^{+    0.011+    0.022}$  & $    0.860_{-    0.011-    0.021}^{+    0.011+    0.022}$  & $    0.8435_{-    0.0089-    0.017}^{+    0.0088+    0.017}$  \\

$H_0 {\rm [km/s/Mpc]}$ & $   71.4_{-    1.6-    3.2}^{+    1.6+    3.3}$ & $   70.1_{-    1.3-    2.4}^{+    1.2+    2.5}$  & $   73.0_{-    1.1-    2.1}^{+    1.1+    2.1}$  & $   71.85_{-    0.98-    1.9}^{+    0.97+    1.9}$  & $   71.11_{-    0.89-    1.8}^{+    0.88+    1.8}$  \\

$N_{\rm eff}$ & $    2.93_{-    0.19-    0.37}^{+    0.18+    0.38}$ &  $    2.82_{-    0.17-    0.34}^{+    0.17+    0.34}$  & $    3.08_{-    0.14-    0.28}^{+    0.14+    0.28}$  & $    3.05_{-    0.14-    0.28}^{+    0.14+    0.29}$  & $    2.87_{-    0.14-    0.26}^{+    0.13+    0.27}$  \\

$S_8$ & $    0.818_{-    0.016-    0.031}^{+    0.016+    0.032}$ &  $    0.829_{-    0.013-    0.026}^{+    0.013+    0.027}$   & $0.813_{-    0.015-    0.030}^{+    0.015+    0.030}$   & $  0.829_{-    0.013-    0.027}^{+    0.014+    0.027}$ & $    0.8111_{-    0.0095-    0.019}^{+    0.0095+    0.018}$  \\

$r_{\rm{drag}}$ & $  148.283_{-    1.867-    3.749}^{+    1.888+ 3.865}$ & $  149.158_{-    1.785-    3.528}^{+    1.806+    3.652}$  & $  146.832_{-    1.472-    2.808}^{+    1.464+    2.922}$ & $  146.906_{-    1.482-    2.942}^{+    1.466+    2.974}$  & $  148.831_{-    1.406-    2.804}^{+    1.352+    2.845}$ \\


\hline\hline                                                                                                                    
\end{tabular}   }                                                                                                                
\caption{Observational constraints at 68\% and 95\% CL using various cosmological datasets within the extended  PEDE+$N_{\rm eff}$ scenario. }
\label{tab:neff}                                                                                                   
\end{table*}                                                                                                                     
\end{center}                                                                                                                    
\endgroup                                                                                                                       

\subsection{PEDE $+M_{\nu}$+$N_{\rm eff}$}

\begin{figure*}
\includegraphics[width=0.5\textwidth]{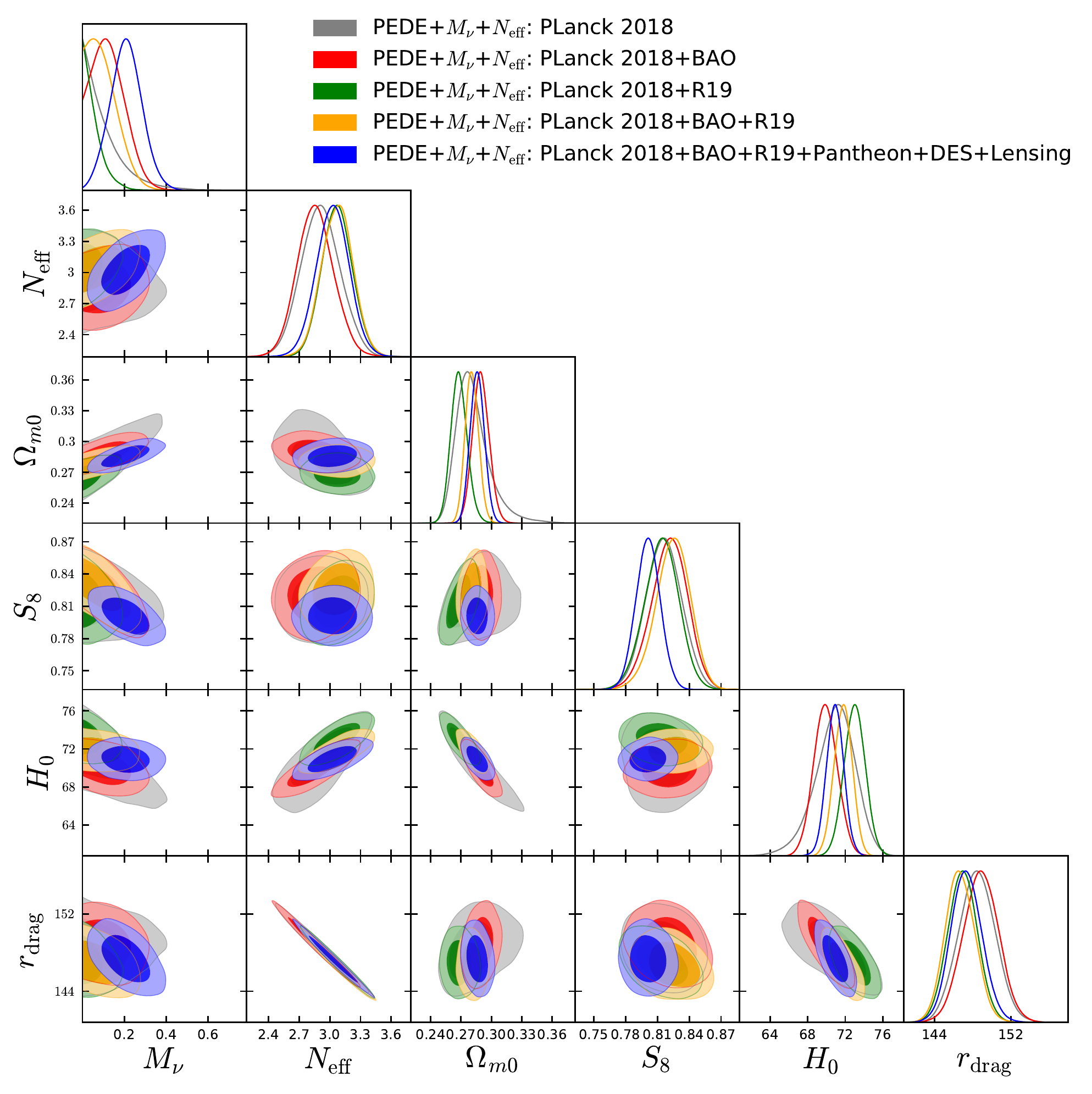}
\caption{One-dimensional marginalized posterior distributions for some selected parameters and two dimensional $68\%$ and $95\%$~CL joint contours for several combinations of the parameters within the extended scenario PEDE + $M_\nu$ + $N_{\rm eff}$ using various cosmological datasets. }
\label{fig3}
\end{figure*}

We report the constraints at 68\% and 95\% CL on the cosmological parameters for  PEDE $+M_{\nu}$+$N_{\rm eff}$ model in Tab.~\ref{tab:mnuNeff} and we show the 1D posterior distributions and 2D correlation plots in Fig.~\ref{fig3}. Comparing Tab.~\ref{tab:mnuNeff} with the previous cases, we find a combination of all the characteristic features we have already discussed in the previous sections, albeit with a larger increase of the error bars.

For the Planck 2018 alone case we find in fact that  $H_0=70.9^{+2.4}_{-1.8}$ m/s/Mpc at 68\% CL, in agreement with R19 at 1.1$\sigma$, $M_{\nu}<0.30$ eV at 95\% CL (slightly larger than the previous cases), and $N_{\rm eff}=2.91\pm 0.19$ at 68\% CL, slightly lower but perfectly consistent with its standard value, $3.045$. For Planck 2018 + BAO, and Planck 2018 + R19 instead, we have exactly the same constraints on the neutrino parameters as we found within the PEDE + $M_{\nu}$ and PEDE + $N_{\rm eff}$ scenarios.

Finally, for the full dataset combination, i.e.~Planck 2018+BAO+Pantheon+R19+DES+lensing, we find that the Hubble constant tension is lowered down to 1.8$\sigma$, a value of the total neutrino mass different from zero well above $3\sigma$ significance ~$M_{\nu}=0.21^{+0.20}_{-0.18}$ eV, and $N_{\rm eff}=3.03\pm0.16$ at 68\% CL, perfectly in agreement with the expected canonical value. The value of $S_8$ is also considerably lower than what one finds within the  $\Lambda$CDM + $M_{\nu}$ + $N_{\rm eff}$ scenario.

\begingroup    

\squeezetable                                                                                                                   
\begin{center}                                                                                                                  
\begin{table*}                                                                         \resizebox{\textwidth}{!}{                                          
\begin{tabular}{ccccccccccccccccc}                                                                                                            
\hline\hline                                                                                                                    
Parameters & Planck 2018 & Planck 2018+BAO & Planck 2018+R19 & Planck 2018+BAO+R19  &  Full \\ \hline
$\Omega_c h^2$ & $    0.1181_{-    0.0032-    0.0057}^{+    0.0029+    0.0060}$ & $    0.1180_{-    0.0032-    0.0057}^{+    0.0028+    0.0061}$ & $    0.1198_{-    0.0028-    0.0055}^{+    0.0028+    0.0055}$  & $    0.1212_{-    0.0027-    0.0053}^{+    0.0027+    0.0052}$  & $    0.1193_{-    0.0026-    0.0053}^{+    0.0026+    0.0051}$  \\

$\Omega_b h^2$ & $    0.02225_{-    0.00023-    0.00046}^{+    0.00022+    0.00043}$ & $    0.02216_{-    0.00019-    0.00037}^{+    0.00019+    0.00037}$ & $    0.02245_{-    0.00017-    0.00033}^{+    0.00017+    0.00033}$  & $    0.02235_{-    0.00016-    0.00031}^{+    0.00016+    0.00032}$  & $    0.02239_{-    0.00016-    0.00033}^{+    0.00016+    0.00032}$ \\

$100\theta_{MC}$ & $    1.04112_{-    0.00044-    0.00087}^{+    0.00045+    0.00088}$ & $    1.04112_{-    0.00044-    0.00087}^{+    0.00045+    0.00090}$   & $    1.04096_{-    0.00042-    0.00081}^{+    0.00042+    0.00082}$  & $    1.04080_{-    0.00041-    0.00080}^{+    0.00041+    0.00082}$   & $    1.04094_{-    0.00044-    0.00078}^{+    0.00039+    0.00086}$  \\

$\tau$ & $    0.0534_{-    0.0081-    0.015}^{+    0.0075+    0.016}$ & $    0.0525_{-    0.0075-    0.015}^{+    0.0075+    0.015}$  & $    0.0550_{-    0.0075-    0.015}^{+    0.0076+    0.015}$   & $    0.0533_{-    0.0076-    0.015}^{+    0.0075+    0.016}$  & $    0.0550_{-    0.0076-    0.015}^{+    0.0075+    0.016}$  \\

$n_s$ & $    0.9598_{-    0.0088-    0.018}^{+    0.0087+    0.017}$ &  $    0.9562_{-    0.0072-    0.014}^{+    0.0073+    0.014}$   & $    0.9674_{-    0.0062-    0.012}^{+    0.0061+    0.012}$  & $    0.9643_{-    0.0061-    0.02}^{+    0.0061+    0.013}$ & $    0.9648_{-    0.0064-    0.0131}^{+    0.0064+    0.0126}$   \\

${\rm{ln}}(10^{10} A_s)$ & $    3.037_{-    0.018-    0.036}^{+    0.018+    0.037}$ &  $    3.035_{-    0.018-    0.036}^{+    0.018+    0.035}$  & $    3.045_{-    0.017-    0.033}^{+    0.017+    0.033}$  & $    3.045_{-    0.017-    0.035}^{+    0.017+    0.036}$ & $    3.043_{-    0.018-    0.034}^{+    0.017+    0.035}$   \\

$\Omega_{m0}$ & $    0.282_{-    0.020-    0.031}^{+    0.011+    0.036}$ &  $    0.2896_{-    0.0081-    0.016}^{+    0.0081+    0.017}$  & $    0.2682_{-    0.0087-    0.015}^{+    0.0075+    0.017}$  & $    0.2807_{-    0.0070-    0.012}^{+    0.0063+    0.013}$  & $    0.2861_{-    0.0069-    0.013}^{+    0.0069+    0.014}$   \\

$\sigma_8$ & $    0.842_{-    0.015-    0.057}^{+    0.030+    0.045}$ &  $    0.836_{-    0.018-    0.042}^{+    0.023+    0.039}$  & $    0.861_{-    0.012-    0.030}^{+    0.016+    0.028}$  & $    0.853_{-    0.015-    0.037}^{+    0.021+    0.034}$   & $    0.821_{-    0.015-    0.030}^{+    0.016+    0.030}$ \\

$H_0 {\rm [km/s/Mpc]}$ & $   70.9_{-    1.8-    4.4}^{+    2.4+    4.2}$ &  $   69.9_{-    1.3-    2.5}^{+    1.2+    2.5}$  & $   73.0_{-    1.1-    2.2}^{+    1.1+    2.2}$   & $   71.77_{-    0.97-    1.9}^{+    0.96+    1.9}$  & $   70.92_{-    0.92-    1.8}^{+    0.90+    1.8}$  \\

$M_\nu {\rm [eV]}$ & $<0.121\,,<0.30$ &  $    0.13_{-    0.10}^{+    0.05} <0.26$ & $ <0.064\,,<0.15$ & $    <0.127 <0.23$ & $    0.209_{-    0.075-    0.14}^{+    0.074+    0.15}$ \\

$N_{\rm eff}$ & $    2.91_{-    0.19-    0.36}^{+    0.19+    0.38}$ & $    2.86_{-    0.19-    0.34}^{+    0.17+    0.35}$ & $    3.07_{-    0.15-    0.28}^{+    0.15+    0.30}$  & $    3.07_{-    0.15-    0.30}^{+    0.15+    0.30}$ & $    3.03_{-    0.16-    0.32}^{+    0.16+    0.32}$ \\

$S_8$ & $    0.816_{-    0.017-    0.033}^{+    0.017+    0.033}$ &  $    0.821_{-    0.016-    0.035}^{+    0.018+    0.032}$  & $    0.814_{-    0.016-    0.033}^{+    0.017+    0.031}$ & $    0.825_{-    0.015-    0.033}^{+    0.017+    0.032}$  & $    0.801_{-    0.011-    0.022}^{+    0.011+    0.023}$ \\

$r_{\rm{drag}}$ & $  148.456_{-    1.897-    3.692}^{+    1.909+    3.741}$ & $  148.842_{-    1.804-    3.606}^{+    1.940+    3.600}$ & $  146.951_{-    1.532-    2.999}^{+    1.520+    3.039}$  & $  146.689_{-    1.630-    2.907}^{+    1.522+    3.045}$  & $  147.352_{-    1.707-    3.092}^{+    1.559+    3.285}$  \\


\hline\hline                                                                                                                    
\end{tabular}   }                                                                                                                
\caption{Observational constraints at 68\% and 95\% CL using various cosmological datasets within the extended  PEDE + $M_\nu$ + $N_{\rm eff}$ scenario.}
\label{tab:mnuNeff}                                                                                                   
\end{table*}                                                                                                                     
\end{center}                                                                                                                    
\endgroup

\begin{table*}[]
    \centering
    \begin{tabular}{cccccccccc}
\hline 
\hline 
      Model & Datasets & $\ln B_{ij}$   &  \\
      \hline 
PEDE        &      CMB       &         $0.6$\\
PEDE        &      CMB+BAO   &         $-1.7$\\
PEDE         &     CMB+R19    &        $7.9$\\
PEDE        &      CMB+BAO+R19  &      $3.7$\\
PEDE          &    CMB+BAO+R19+Pantheon+DES+Lensing   &  $-3.1$\\
\hline
 
PEDE + $M_{\nu}$      &    CMB        &        $0.1$\\
PEDE + $M_{\nu}$      &    CMB+BAO     &       $-1.0$\\
PEDE + $M_{\nu}$      &    CMB+R19       &     $6.9$\\
PEDE + $M_{\nu}$      &    CMB+BAO+R19    &    $3.7$\\
PEDE + $M_{\nu}$      &    CMB+BAO+R19+Pantheon+DES+Lensing   &  $-1.5$\\
\hline 
PEDE + $N_{\rm eff}$     &    CMB        &        $0.2$\\
PEDE + $N_{\rm eff}$     &    CMB+BAO      &      $-2.5$\\
PEDE + $N_{\rm eff}$     &    CMB+R19       &     $4.1$\\
PEDE + $N_{\rm eff}$     &   CMB+BAO+R19    &     $3.2$\\
PEDE + $N_{\rm eff}$     &   CMB+BAO+R19+Pantheon+DES+Lensing  &   $-3.1$\\
\hline 

PEDE + $M_{\nu}$ + $N_{\rm eff}$   &  CMB      &         $0.3$\\
PEDE + $M_{\nu}$ + $N_{\rm eff}$   &  CMB+BAO   &        $-0.9$\\
PEDE + $M_{\nu}$ + $N_{\rm eff}$   &  CMB+R19     &      $4.5$\\
PEDE + $M_{\nu}$ + $N_{\rm eff}$   & CMB+BAO+R19   &     $3.4$\\
PEDE + $M_{\nu}$ + $N_{\rm eff}$   & CMB+BAO+R19+Pantheon+DES+Lensing   &              $-1.2$\\
\hline \hline 
\end{tabular}
\caption{Summary of the $\ln B_{ij}$ values quantifying the evidence of fit of the cosmological scenarios with respect to a base cosmological model $\Lambda$CDM considering the observational datasets employed in this work. We note that the negative value of $\ln B_{ij}$ indicates that $\Lambda$CDM is preferred over the model. }
    \label{tab:BE}
\end{table*}

\section{Summary and Conclusions}
\label{sec-conclu}

Our work has been mostly motivated by the measurement of one of the key cosmological parameters -- the Hubble constant ($H_0$) -- which has been a serious issue at present time demanding alternatives to the well known $\Lambda$CDM cosmology. A few years back, it was noticed that the estimations of $H_0$ by two different observational missions did not really match at all. This mismatch is currently highly significant: the recent observational results on this issue report that the estimated value of the Hubble constant by Planck~\cite{Aghanim:2018eyx} within the minimal $\Lambda$CDM paradigm is more than $\sim 4 \sigma$ apart from the estimation by Riess et al (SH0ES collaboration)~\cite{Riess:2019cxk}.
While a possible pure systematic origin of the tension~\cite{Efstathiou:2013via} is still under debate, a number of other possible avenues have been followed in the literature. 
An independent high versus low redshift discrepancy is that related to the substantial discordances among CMB measurements from Planck and those from cosmic shear  concerning the matter perturbations at small scales. The tension is usually  quantified  in  terms  of the $S_8$ parameter. To alleviate these tensions, especially the Hubble constant one, some approaches have been followed (dynamical or interacting dark energy, early dark energy,  modified gravity and other non-standard scenarios). Despite the large differences among these cosmologies, all of them share a common feature, and it is the fact that mostly all include an extension of the six parameter-$\Lambda$CDM cosmology.  

Here instead we focus on a phenomenological alternative which does not imply any extra degrees of freedom: the Phenomenological Emergent Dark Energy model (PEDE), a possible \emph{minimal} scenario which has been shown to provide an excellent solution to the long-standing $H_0$ tension. Also the values of the $S_8$ parameter are lower than in the canonical $\Lambda$CDM cosmology and, therefore, {\it the model is a promising avenue towards solving the current cosmological tensions.} 
However, before establishing the PEDE model as a concrete and complete framework which could serve as a guidance for model building, it is mandatory to ensure that the resolution of the discrepancies is stable against obvious extensions of the PEDE model. We have therefore re-analyzed the PEDE scenario with the inclusion of massive neutrino species and possible extra relativistic degrees of freedom, because they can help with the sound horizon problem.
We find that the cosmological tensions are still alleviated: the Hubble constant tension is always below the $2\sigma$ level and the value of $S_8$ is in a better agreement with cosmic shear estimates. With the most complete combination of datasets, we have obtained $M_{\nu}\sim 0.21^{+0.15}_{-0.14}$~eV  and $N_{\rm eff}= 3.03\pm 0.32$ (with $95\%$~CL errors), i.e. an indication for a non-zero neutrino mass with a significance above $2\sigma$. However, the sound horizon is still $\sim 147$ Mpc, failing in recovering the lower value preferred by the BAO data. 
While the sound horizon issue is not solved in this case but based on other outcomes in this article, one may believe that the PEDE model provides a new alternative minimal and solid framework to inspire other phenomenological possibilities. This feature is strengthened through the Bayesian Evidence analyses (see Table \ref{tab:BE} where we present the values of $\ln B_{ij}$ for PEDE and its several extensions) which clearly show that for CMB, CMB+R19 and CMB+BAO+R19, PEDE and its all extensions are preferred over $\Lambda$CDM whilst for the remaining two datasets, namely, CMB+BAO and CMB+BAO+R19+Pantheon+DES+Lensing, $\Lambda$CDM is favored. If future laboratory measurements will find an indication for $M_{\nu}\sim 0.21$~eV, this would strongly motivate phenomenological PEDE  
like-scenarios, as alternatives to the $\Lambda$CDM canonical picture.

\bigskip 

\begin{acknowledgments}
The authors thank the referee for some clarifying points which completed the article. 
WY was  supported by the 
National Natural Science Foundation of China under Grants 
No. 11705079 and No. 11647153. 
EDV was supported from the European Research Council in the form of a Consolidator Grant 
with number 681431. SP has been supported by the Mathematical Research Impact-Centric Support Scheme 
(MATRICS), File No. MTR/2018/000940, given by the Science and Engineering Research Board 
(SERB), Govt. of India. O.M. is supported by the Spanish grants FPA2017-85985-P, PROMETEO/2019/083 and by the European Union Horizon 2020 research and innovation program (grant agreements No. 690575 and 67489).
\end{acknowledgments}

\bibliography{biblio}
\end{document}